\documentstyle[12pt]{article}
\begin{document}
\begin{center}
{\Large Quantum creation and inflationary universes: a critical
appraisal }\\
\bigskip
{\bf D.H. Coule}\\
\bigskip
School of Mathematics\\ University of Portsmouth, Mercantile
House\\ Hampshire Terrace, Portsmouth PO1 2EG
\bigskip
\begin{abstract}
We contrast the possibility of inflation starting a) from the
universe's inception or b) from an earlier non-inflationary state.
Neither case is ideal since a) assumes quantum mechanical
reasoning is straightforwardly applicable to the early universe;
while case b) requires that a singularity still be present.
Further, in agreement with Vachaspati and Trodden [1] case b) can
only solve the horizon problem if the non-inflationary phase has
equation of state $\gamma<4/3$, so excluding radiation or massless
scalar field dominated cases. Other alternative models, such as
the smooth branch change in the pre-big bang model,  have related
problems of requiring homogeneity over non-causally connected
large scales.

\end{abstract}


\end{center}
\newpage
{\bf 1 Introduction}

The standard big bang model (SBB) model has a number of puzzles,
particulary the so-called horizon and flatness problems, that are
believed to  require  explanation and not just be accepted by
fiat, see eg.[2,3]. However, for matter that obeys the strong
energy condition, so including radiation or dust sources, there is
by necessity an initial singularity, see for example [4]. On such
a singular surface one can impose any suitable data one wants to
give rise to our universe after it evolves into the future [5,6].
So strictly speaking the presence of a singularity obviates any
talk of being unnatural since what is natural at a singular
surface. A somewhat related argument has been made that at the
singular surface everything can naturally becomes smooth by
entropy arguments[7].

The more generally accepted solution to solving the problems of
the SBB is to allow a matter source that violates the strong
energy condition and which can give rise to a rapidly expanding
regime, so called inflation e.g. [2,3]. One imagines that a small
homogeneous initial patch can be taken and massively expanded to
give our present universe. However, a lower bound on the size of
this patch has been made by Vachaspati and Trodden (VT) and which
could actually prevent the horizon problem being solved by
inflation [1]. We will consider these claims  in detail.

Throughout this article we will take so-called chaotic inflation
as the prototype for inflation [2,3] . It can be caused by a
scalar field with potential $V(\phi)$ sufficiently displaced from
its minimum. However, one still needs to know the distribution for
the initial conditions of this scalar field. One natural measure
for a classical system is the canonical one [8]. But using this
measure, and even after assuming an homogeneous scalar field, the
probability for inflation is found to be ambiguous: an infinity of
both inflationary and non-inflationary solutions [9]. Further, the
predicted  ${\em flatness}$ becomes arbitrary if there is an upper
bound in the potential $V(\phi)$, coming from perhaps a conformal
anomaly correction or from a higher derivative $R^2$ term in the
action [10].

 One obvious limitation to these arguments is that the
classical equations will fail as the quantum gravity epoch is
encountered near the initial singularity. The initial conditions
should then be supplied by a suitable quantum measure. One
approach using the Wheeler-DeWitt (WDW) equation aims to achieve
this, see eg. [2,3]. Although different results are obtained
depending upon which boundary conditions are believed correct. The
two most common approaches are the ``no boundary''of
Hartle-Hawking [11] and Tunnelling ones [12]. There is some
disagreement as to whether inflation is actually predicted with
the Hartle-Hawking boundary condition [13,14], but the Tunneling
one seems more certain . We just note that the quantum measure can
give a more certain prediction compared to a purely classical
approach because the initial size of the universe is necessarily
``small'' when quantum mechanics is relevant. An initial ``quantum
of energy'' can be distributed amongst the kinetic and potential
energy of the scalar field. If the boundary conditions allow the
potential to dominate then a rapid inflationary expansion ensues.
Roughly speaking, large positive energy is produced by producing
large volumes with negative gravitational energy, which can
subsequently create large amounts of matter. The possibility of
inflation is of course dependent on there being present a suitable
scalar potential. This is put in by hand and not a prediction of
the quantum cosmological theory. The actual source of this
potential or more generally a reason for the initial violation of
the strong energy condition needs to be understood from the actual
quantum gravitational theory once known, perhaps eventually from
string theory. Until this time inflation cannot be thought of as a
true explanation of the universe's existence but rather a possible
mechanism for amplifying a quantum created universe that is
unnaturally small. Since the creation mechanism is not known it
could just be the case that this mechanism itself could give a
reason for explaining the various puzzles and allowing a larger
initial universe. Recall the present visible  universe
extrapolated back to the Planck time is a size $\sim 10^{-3}cm$, a
factor $\sim 10^{30}$ larger than the Planck length $\sim
10^{-33}cm$.  In other words the creation mechanism is assumed to
be driven by quantum effects that are more probable only over
small scales, but since notions of time and energy  are only
properties within our universe it isn't clear whether quantum
mechanics constrains the initial creation in such a way.

 {\bf 2 Inflation starting from inception }

Given, that some quantum process does lay down the initial
conditions one can try to justify the likelihood of inflation
being present. This is usually done at the Planck epoch where the
semiclassical equations are first valid although as mentioned this
should only be taken as a rough first guess . For example, in
string theory recent developments suggest the Planck scale in
higher dimensions could be at a lower energy scale [15]. Although
from experimental constraints this cannot account entirely for the
factor $\sim 10^{30}$ that could  alone rectify the unnaturalness
of the SBB model.

 In the case of chaotic inflation the Friedmann equation for
a homogeneous FRW universe is [2,3]
 \begin{equation}
 H^2+\frac{k}{a^2} = \dot{\phi}^2+V(\phi)
 \end{equation}
where we set Newton's constant $8G/3=1$. If the initial Planck
energy is distributed between the kinetic and potential terms of
the scalar field then,
\begin{equation}
\dot{\phi}^2+V(\phi)\leq M^4_{pl}
\end{equation}
with $M_{pl}$ the Planck mass. This constraint is not too severe
for the possibility of obtaining a sufficient potential component
since if the kinetic energy initially dominates the scalar field
only decays with a logarithmic dependence $\phi\propto \ln (t)$
[16-18].

 There is also the question of whether the scalar field is homogeneous
initially. Such a possibility is described by the spatial gradient
terms. The question is not quite the same as the horizon problem
which we will elaborate on later.  For inflation to occur the
potential should dominate over such terms, that is
$V(\phi)>>(\nabla \phi)^2$. Otherwise, such spatial gradients do
not drive inflationary expansion but rather an expansion $a\propto
t$, which is only  on the verge of power-law inflation see
eg.[19]. Provided such gradient terms are not too large they
typically die away faster than the potential energy. Numerical
studies, suggest the inflationary regime will take over provided
the curvature is not too large that the universe could re-collapse
before entering an inflationary phase [19].

The requirement of sub-dominance of the spatial gradient term does
at first hint at a problem with causality. Consider a region over
which the initial $\phi_i$ varies a unit amount to be $L$. Then
[3,19]
\begin{equation}
V(\phi)>(\phi/L)^2 \;\;\Rightarrow L>\frac{\phi_i}{V^{1/2}}
\end{equation}
Now because in a flat universe $H^2=V(\phi)$ this constraint
suggests that the field $\phi$ is smooth over scales $L>\phi
H^{-1}$. Since an initial field $\phi_i>>1$ is required for
sufficient inflation the smoothing scale is larger than the
``causal horizon'' $\sim H^{-1}$. For a quadratic potential
$V(\phi)=\lambda \phi^4$ the initial field can take values up to
the Planck epoch values  $\sim \lambda^{-1/4} M_{pl}$. The value
of $\lambda$ is extremely small $\sim 10^{-12}$ in order to
suppress quantum fluctuations and give agreement with the COBE
measurements [2,3]. This gives that the corresponding  smoothing
scale or wavelength of the inhomogeneity should  be $\sim 10^4
H^{-1}$.
 However, since the initial conditions, of some finite starting
  value for $H$, are simply being imposed
by some Planck value constraint this requirement has nothing {\em
per se} to do with causality. The properties of the universe are
assumed to be given by some global creation process and the
behaviour of light rays, that define the causality constraints,
within such a domain are immaterial. The large smoothing scale
does still require some justification from a theory of initial
conditions. If the initial domain was instead assumed to be
chaotic then a smoothing process would need to be postulated in
order to make conditions suitable for inflation to occur. Such a
mechanism would then, at least classically, be expected to work
only over causal scales.

We note in passing that in assisted inflation the smoothing scale
might easier be explained [20] . In such models    a number of
small field, which each individually would not produce inflation,
might act in concert to do so. Although, in chaotic assisted
inflationary models with $\phi^n$ potentials they seem susceptible
to cross coupling terms; they further assume the fundamental
Planck scale coming from a higher dimensional theory is at a lower
energy scale than the usual 4 dimensional one [21]. If the field
was naturally smooth over length scales  approximately like the
higher dimensional Planck length it could naturally appear smooth
over length scales much bigger than the  4-dimensional Planck
length, when dimensional reduction occurs.

 If the initial domain is not created uniformly then the
 homogeneous patch that is to inflate must be of
 sufficient size. Recall that the scalar field behaves as a
 negative pressure and any outside positive or zero pressure will
 wish to equalize the situation by rushing in. Assuming this
 equalization can proceed at the speed of light one finds that the
 homogeneous domain must be of a size greater than $\sim 3H^{-1}$ [19].
 One can speculate that other ways of isolating the interior from higher
 pressure and weakening this requirement are possible
  cf.[22] where large negative
 pressure of water in various natural systems can occur. In the
 inflationary case this could correspond to a highly
 non-trivially topological structure where a negative pressure
 might be  isolated from its surroundings.
 Once such suitable domains occur then inflation will proceed.
 Because of a combination of  the fluctuations
 that are generated during inflation and
 the finite horizon size $\sim H^{-1}$ in De Sitter
 space inflation never stops entirely once started, provided a
 requirement on the size of the scalar field $\phi>\phi_*$ is achieved [2].
  If the initial domain
 does not have a sufficiently large scalar field value one can hope that
 a quantum ``instanton'' effect can produce a domain with sufficiently large
 scalar field provided there is sufficient starting
  volume that the probability
 is not infinitesimally small [2,23].  One can try
 and make inflation eternal into the past but geodesic completeness
  is not possible without violating the more extreme weak energy
  condition [24]. A singularity is therefore present if one tries to extend
  an inflationary phase backwards indefinitely. Without some
  creation mechanism to give the initial inflationary conditions the model
  would still suffer the same initial singularity problem as in
  the usual SBB model. The quantum fluctuations also drive the
  average field into Planck energy densities where the theory is
  unknown, although one can argue that inflation is still future
  eternal even if an absorbing boundary is imposed at the Planck
  scale to remove such singular states[23]. One worrying aspect of
  inflation is that a non-minimal coupling i.e. $\xi R \phi^2$,
  can,
  apart from preventing inflation entirely for larger $\xi$,
  change the renormalization schemes that are required to produce
  a linear growth in $<\phi^2>$ that is required to drive eternal
  inflation cf. [25]. The large fluctuations might also cause black hole
  or other defect
  formation, which will reduce the surface gravity and so possibly
  regulate the eternal inflationary mechanism cf.[26,27] .

  {\bf 3 Quantum creation of initial state }

  How realistic is the concept of quantum creation ? We would just
  like to mention a few aspects of this problem that often seem
  lost among all the technicalities in the literature.

In usual quantum mechanics one is given the relevant potential, or
action in the path integral formulism, prior to doing a
calculation. The situation is more confused in quantum cosmology
since the potential, determined partly by the matter components,
is supposedly  also being predicted. One might wonder if this is
at all sensible, or at least that the notion of ``creation'' is
overstated since the relevant quantities are assumed to exist
prior to starting the calculation. The formulism more accurately
tries to obtain which quantities are most dominant amongst all the
allowed possibilities: ``quantum determination of the cosmological
state'' would be a more apt description.

With this more limited aim let us consider further the
  ``quantum tunneling from nothing'' and
  Hartle-Hawking (HH) schemes. They both depend crucially
  on the WDW potential, given, in the
  simplest models by [2,3]
  \begin{equation}
  U=ka^2- V(\phi)a^4
  \end{equation}
  being positive to produce a forbidden or Euclidean region. But
  this requires a number of assumptions:

  i) The curvature $k$ must be closed ($k=1$) to produce a
   forbidden region ($U>0$) for $a$ small. This region
  can then be  tunnelling through, or oppositely  ``anti-tunnelled''
    in the HH case. But
  this naive treatment of the curvature is precisely the quantity
  that is likely to be modified as general relativity is
  superceded by quantum gravitational effects. Because with
  $V(\phi)$ roughly constant, the
  strong energy condition is being violated curvature dominates at
  small scales unlike for usual matter (eg. radiation)
   where the curvature only becomes
  relevant at late times when the scale factor is large. One might
  expect curvature to wildly vary on  small scales and the crucial
  $ka^2$ term to be inadequate for its full description.

  In flat
  or open geometries there is no forbidden sector,
  but the model can still be quantized although the relevant
  boundary conditions to impose are less well motivated [28]. Assuming
open compact geometries one can argue that inflation is likely by
imposing analogous ``outgoing like'' boundary conditions[28].
There is a further possibility of removing forbidden regions by
use of signature change [29]. Because the lapse function is
arbitrary and not determined by the Einstein field equations it
can change sign to prevent the forbidden region being classically
disallowed. With such a signature change  variable present, one
can further quantize the model, although there seems further scope
for different boundary conditions [30-32]. These extra
possibilities seems symptomatic of what could occur as the Planck
scale is approached and curvature and time become less structured
cf.[33].
\\

  ii) Matter that satisfies the strong energy condition also has to be
  removed to create a Euclidean region at small size.
  This is done indirectly by imposing boundary
  conditions that disallow such matter
  to dominate as $a\rightarrow 0$. As emphasized by ref.[34] quantum
  uncertainty should at
  least introduce such a matter source of ``zero point''
  size. But any such matter eg. radiation, stiff matter, now dominates
  over the curvature or $V(\phi)$ term. For stiff
  matter or equivalently the kinetic term of a scalar field a
  singularity represented by a wildly oscillating
   wave function is produced.
  Even if the singularity is absent or suitably
  regularized ( possibly by using
  a separation constant-see eg.[28]) the
  model represents a Lorentzian model similar
   again to  the case with non-closed
  curvature present. The constants that determine the matter
  content of this universe are arbitrary, just as the size of an
  hydrogen atom is determined by the potential and not the
  quantization {\em per se}. In these cases the universe can have
  sufficient size that inflation would be unnecessary and
  depending upon the initial perturbation spectrum could give a
  suitable model for our universe.
  \\

  iii) When strong energy satisfying matter is present eg.
  radiation, one typically
  needs to impose the boundary condition $\Psi(a=0)
  =0$ [34]. But this
   certainly does not explain why the universe comes into
  existence and indeed is not consistent if one considers zero
  scale factor to be the relevant starting value. Likewise the inclusion
  of  a forbidden region isolating $a=0$ from the rest of the universe
  does not explain why the zero scale factor should have special status
  and not itself require explanation. Quantum tunneling is from
  one well defined state to another: it is not a magical process
  that explains something from nothing. The actual cosmological problem
is further compounded by the fact the potential itself is being
predicted.
\\

iv)  One can even question whether the aim of obtaining the
initial
  value of $V(\phi)$, to determine the ensuing amount
   of inflation, is achievable. Consider the analogous problem of
  field emission or enhancing the alpha particle decay of a nucleus by
  applying an electric field $E$ - see eg.[36]. The relevant  potential is
  \begin{equation}
  U=W-Ea
  \end{equation}
  where $W$ is the work function of the metal in the field emission
  problem, but which for our purpose is simply a constant.
   Note how in the quantum
  cosmology case the potential $V(\phi)$ is analogous to the
  electric field ( ignoring some unimportant discrepancy in the factors of
  $a$). Now in the field emission problem, one asks: given an
  applied electric field what is the probability of electron
  emission? The answer is proportional to $\sim \exp(-1/E)$, so
  a larger electric field enhances the emission. But in the
  quantum cosmology problem we are trying to ask: given tunneling
  occurred ( a universe exists) what
   is the value of the potential $V(\phi)$?  In the
  electron case this would be indeterminate for a single event,
  but given there is a reservoir of electrons that all eventually
  tunnel this problem is still solvable, by repeated observation,
   given that every other variable
  of the problem is known. The longer the  observation the more
  certain the applied field can known.
   But for the alpha decay which is also a
  single one off event, there is an ambiguity in trying to know the
  applied field if you don't know how long the  system has already remained
  undecayed. And the quantum cosmology case is, at least as
  complex, and
   analogous to this
  problem since the time being absent means there is no time limit on
  when the decay (or initial creation) should proceed. One might
  try to argue that if $V(\phi)$ is initially
  too small the Euclidean region
  would extend over large scale factors, and similarly for
  negative $V(\phi)$ the universe is entirely Euclidean. Indeed, the HH
  scheme apparently favours the Euclidean
   region extending to large scales, while the
   ``tunneling from nothing'' prefers it
   confined to small scales [37].  Without
  use of anthropic arguments one might have to postulate a ``source
  of universes''  that can come into existence, analogous to the
  electrons in the metal, so that ``on average'' $V(\phi)$ can
  be assumed to be large. But because time is not even existing
  exterior to the universes even this sort of modification  is
  interpretatively suspect, since how would the ``reservoir'' be
  defined. Requiring a surfeit of universes is also rather
  extravagant.  See also ref.[38] in this regard, who try to suggest
  why universes are not still being continuously created around
  us, but this rapidly starts getting into the quagmire  of different
  interpretations of quantum mechanics. The points I have made are
  rather general and mostly independent of more abstruse arguments
  about which interpretation of quantum mechanics is valid,
  although ultimately this could have a bearing. Working with the
  path integral formulism would raise equivalent problems and  concerns.

In summary, the arguments of the ``no boundary'' or
   ``tunneling from nothing'' proposal, which respectively are claimed to
    give distributions $\sim \exp (1/V(\phi))$ and
   $\sim \exp(-1/V(\phi))$ [2,3] for the initial $V(\phi)$,
    are at best provisional,
   with the usual quantum analogies and
    definitions of probability applied in a  cavalier manner. One
    at least hopes that quantum mechanics can still
     be reasonably applied and is actually valid during the early epoch.

Returning, again to some general notions of inflation.
 It is often supposed that the scale invariant
fluctuations that are generated during inflation and the
subsequent Doppler peaks in the CMB spectrum are a validation of
inflation. But as emphasized in ref.[6] such a signature does not
validate inflation since other models with fluctuations created
early  can give  such behaviour [6], and of sufficient size [39] .
An example of an alternate mechanism for fluctuations  is if a
state of ``self-organized criticality'' could occur  during the
early universe. So giving fluctuations over scales of many
magnitudes [40]. Such processes, which are believed a general
feature of nature [41], could occur during phase transitions at
the Planck scale, hopefully without requiring special initial
conditions.

Because  of our limited knowledge of quantum creation  it is
difficult to know whether a subsequent inflationary stage will be
necessary. Instead of treating inflation as an additional phase to
correct the failings of the initial creation process one might
just as well, with our present understanding, include it within
the initial creation scheme. Such a scheme has the advantage of
regularizing singularities, although it still depends on the type
of  matter  present and factor ordering terms. But where the
singularities are present is precisely when the quantum gravity
corrections are most important and the WDW equation as presently
formulated is unlikely to be  valid. In summary, the initial state
resulting from a creation scheme is rather provisional. Whether it
requires a subsequent inflationary stage is actually  unknown
until this state is better defined.

 {\bf 4 Inflation from
previously non-inflationary conditions}

 What about the naturalness of inflation in models that don't
inflate initially from their inception.
  Singularities seem a general consequence of producing conditions
  that give inflationary behaviour. The idea of producing a
  ``universe in the lab'' was required to expand so rapidly to
  avoid re-collapse that a singularity would be present [42]. This can
  be seen clearer by noting that in a FRW universe regions of size
  bigger than the so-called apparent horizon $\sim \rho^{-1/2}$
   have a necessary singularity -see page 353 in ref.[4]
   where such a quantity
   is called the {\em Schwarzschild length} of matter density $\rho$.
  But this size, $\sim H^{-1}$ for
   the flat $k=0$ case, is the minimum required to isolate an inflationary
  patch from its surroundings for sufficient time to start
  inflating. In fact requiring the initial patch  size to be
  larger than the apparent horizon size has recently been claimed
  by VT
  to be a major problem in setting up inflationary conditions within
  a patch. However, depending on the matter source it need not strictly
  violate causality which is rather determined by the particle horizon: the
  distance light travels from the beginning of the universe. The
  large initial patch size means rather that a singularity is present.

Let us first recall  the nature of the horizon problem. It
 occurs because the the particle horizon size, defined as
\begin{equation}
r=c\int_{0}^{t} \frac {dt}{a(t)}
\end{equation}
is finite, see eg.[2,3]. The horizon proper distance $R$ is
 this quantity $r$
multiplied by the scale factor i.e. $R=a*r$. For any strong-energy
satisfying matter source this quantity $R$ grows linearly with
time. But in SSB cosmology the rate of change of the scale factor,
given by $a\sim t^p$ and $1/3<p<1$, grows increasingly rapidly as
$t\rightarrow 0$. The horizon cannot keep pace with the scale
factor `velocity' $\dot{a}\sim 1/t^{1-p}$. But note that this is
only impossible for times below unity $0<t<1$. If the horizon
problem was solved, by some (quantum)  process, at the Planck time
$t_{pl}=t= 1$
 it would remain  permanently solved during the ensuing evolution [43].
  Note
 also that in models that inflate from their inception the usual
space-like singularity of the FRW universe becomes  null like when
$p>1$- see eg.[44]. The idea of inflation is to take an initial
domain of size less than the corresponding particle horizon size
and allow it to expand greatly to encompass our universe. Let us
see how this  requirement can be constrained. Fortunately, most of
the relevant quantities have already been obtained in work on the
holography conjecture and can readily be applied in this context
[45,46].
 For this purpose a  useful form
of the FRW  metric is
 \begin{equation}
 ds^2=a^2(\eta)\left ( -d\eta^2 +d\chi^2 + f^2(\chi) d \Omega^2 \right )
 \end{equation}
 where $f(\chi)=\sinh \chi\; , \chi\; , \sin\chi $ , corresponding
 to open,flat and closed universes
 respectively. We can define a number of important quantities.
 The {\em Hubble horizon} is defined by
 \begin{equation}
 r_H=H^{-1}
 \end{equation}
  The {\em particle horizon}, or the distance travelled by
 light from the initial moment of the universe, is simply,
 \begin{equation}
 \chi_{PH}=\eta
 \end{equation}
 for this metric. The {\em apparent horizon} is given by [45]
 \begin{equation}
 \chi_{AH}=\frac{1}{\sqrt{H^2+k/a^2}}\Rightarrow
 \frac{1}{\sqrt{\rho}}
 \end{equation}
Roughly speaking light rays beyond the apparent horizon are seen
to move away from the origin, a so-called anti-trapped behaviour.
Note that in the flat case $k=0$ the apparent horizon and Hubble
horizon coincide.

 Vachaspati and Trodden [1] have argued that the initial inflationary
patch must have sufficient size $x$ that it reaches the
anti-trapped surface i.e. $x>r\chi_{AH}$. Otherwise the weak
energy condition is violated for light rays that could otherwise
enter the inflating region from normal or trapped regions. For a
perfect fluid with equation of state $p=(\gamma-1)\rho$ the
apparent horizon has the following time dependence [45,46]
\begin{equation}
\chi_{AH}= \frac{d\gamma-2}{2} \eta
\end{equation}
with $d$ the number of space dimension i.e.  3 in the usual 4
dimensional space-time case. However the causal particle horizon
has a different time dependence simply $\chi_{PH}=\eta$ so the
condition
\begin{equation}
\chi_{AH}<x<\chi_{PH}
\end{equation}
can be satisfied for
\begin{equation}
\frac{d\gamma-2}{2}<1 \;\; \stackrel{d=3}{\rightarrow} \;\;
\gamma< 4/3
\end{equation}
This does exclude the case of radiation ($\gamma=4/3$) or stiffer
equations of state. But if $\gamma$ was gradually reducing before
inflation occurred, recall the strong energy condition is violated
as $\gamma$ falls below $2/3$, this causal constraint can be
satisfied. The condition can be thought of as saying the effective
value of $\gamma$ cannot switch suddenly but rather must fall
below $4/3$ for sufficient time to allow the causal or particle
horizon to be larger than the apparent horizon. This result is
independent of whether curvature is present. Although in the open
and flat cases one might argue that the presence of an infinite
spacial section gives a more likely chance that inflationary
conditions would occur to give an inflationary patch.  In the
closed case only during the expansion phase is an anti-trapped
surface present- see Fig.(4) in ref.[46]. This means that
producing inflation to avoid an impending ``big crunch''
singularity during a collapsing phase will violate not only the
strong energy condition.
 Now it is true that needing $x> \chi_{AH}$ is difficult to
justify in terms of particle physics processes, but if this patch
could be smaller than $\chi_{AH}$ one could avoid the singularity
in a FRW universe since the matter would be insufficient to
converge the light rays into the past. See chapter 10 in ref.[4]
for a proof of this argument. So allowing an initial domain of
size $x<\chi_{AH}$ to inflate, would have allowed singularities to
be expunged from this cosmology: the result that this cannot be
done without violating the weak energy condition is therefore
consistent with the studies of eternal inflation that
singularities have to be present when the model is continued into
the past [24].

There are some alternative metrics with non-singular solutions,
but like Minkowski space they don't have anti-trapped regions
[44,47]. Achieving inflation is such spaces would likewise require
the violation of the weak-energy condition.

{\bf 5 Higher dimensions and Brane inflation}

 We can make some remarks about specific inflationary models. For
 example in higher dimensional theories it will be more difficult
 to satisfy the constraint. In 5 dimensions the constraint becomes
 $\gamma<1$ so now excluding the pressureless dust equation of state.
 In 7 dimensions or more the strong-energy condition has to be violated
 from the start to ensure the constraint is met.

 In 5 dimensional Brane models the
 Hubble parameter is further modified, typically such that $H=\rho$, on the
 4 dimensional Brane corresponding to our universe [48]. Matter is
 effectively stiffened and the particle horizon can extend beyond the
 apparent horizon now only for $\gamma<2/3$: which would violates the
 strong energy condition in usual 4-dimensional general relativity.
  Indeed it is known that the modified
 Friedmann equation has a more stringent requirement, that
 $\gamma<1/3$, for inflationary behaviour [48]. But it also appears
 impossible to solve the horizon problem without matter that is
 initially  inflationary anyway. Unless such matter is
  present in string theory, the inflationary conditions would need to be
 understood by non-causal initial conditions or weak energy
 violating quantum mechanisms. Otherwise one again would have to
 postulate inflation from its initial inception.

 There is another approach where the presence of extra branes
 might allow a ``short circuit'' of space-like separations compared to
 when only a single brane is present[49]. This is rather similar to earlier suggestions that
 wormholes could alleviate the horizon problem during an early
 quantum gravitational phase [50]. For such a scheme to work the
 branes have to be curved and aligned in just the right fortuitous manner.

{\bf 6 Pre-Big Bang models}

 The VT requirement is  also relevant for the pre-big bang
 scenario [51]. In this model the universe starts at time minus infinity and
 expands towards a singularity at time zero, which will corresponds to
 the usual Big bang beginning of the universe. I have earlier
 criticized whether this is a true inflationary behaviour but will ignore this
 aspect in the present discussion [52].

    The expansion is driven
 by the dilaton present in string theory. In the Einstein frame this dilaton
 is simply a massless scalar field or equivalently a stiff fluid driven
 model. A mechanism is required to end the contraction phase and
 branch now into  a suitably  expanding behaviour. One assumes that
 higher order corrections become increasingly important as the
 universe becomes increasingly hotter and turbulent.  Because there are
 no anti-trapped regions in a collapsing universe one cannot enter an
 inflationary phase without violating the VT constraint.

  Even for a patch
 just locally
 trying to avoid the singularity and enter a non-inflationary expansion
 there is a similar constraint. As the universe collapses
  one needs to ensure that the particle horizon (starting from the
  time new string states are being produced)
 keeps the universe
  causally connected. In a collapsing model the usual particle horizon
  of an expanding stiff fluid  is
  converted to an event horizon as the singularity is approached. But
  because the matter is constantly being modified by new processes becoming
  important an effective particle horizon is also present.    To avoid the
 singularity and create a smooth post big bang phase
 one needs to set up uniform conditions over a length scale larger than the
 trapped surface. Only for perfect fluids with $\gamma <4/3$ can the
 event horizon remain larger than the size of the trapped surface and one
 might justify an homogeneous distribution of strong energy violating material to avoid the
 impending singularity. In the dilaton driven case it seem that the VT
 constraint will mean that the weak energy condition is necessarily
 violated. Similar conclusions have been obtained in specific models [53]
  but
 the requirement of requiring weak energy condition violation
  is more transparent when using the VT constraint. One might hope that
  the stiff equation of state
  is modified to that below $\gamma=4/3$ before one
  tries to implement the branch change in order to require only
   the strong energy condition to be violated. There is another
   potential problem even if this matter can be distributed. The
   measure for such branch changes or bounces
    is likely to depend on any upper bound
   in some effective scalar potential and singular solutions are
   still likely to be more probable cf.[10]. Such criticism could
   also be levelled at the ``evolution of universes'' scheme of Smolin [54],
    where every collapsing
   black hole is envisioned as bouncing into to another universe.

    To summarize the difficulty, in the
  pre-big bang phase one is sailing too close to the singularity where
  processes occur increasingly  rapidly. Near
  such a surface, horizon problems again result since
   one wishes to influence a
  sufficiently large patch to avoid the singularity and keep the resulting
  post big bang universe smooth and coherent. The situation required
  is rather opposite to the
   usual ending of inflation where parts of the universe
  often remain in the inflationary
   state. In this case none of the universe can
  be allowed to remain pre-big bang
  or inflationary as a space like singularity would
  result. Whether, one
  can justify such a simultaneous  exit therefore depends on knowing what
  quantum gravitational effects will come into play and what
  energy conditions can be justifiably violated. Otherwise
   the pre-big bang model simply  brushes its problems under
  the branch change carpet.

{\bf 7  Variable constant models}

 Another possible solution to the horizon problem is to
 postulate that the various constants particularly $c$ could take
 different values during the early universe [55-57]. This alone is
 not too helpful since a space-like singularity cannot be crossed
 by any finite value of $c$ and a higher $c$ just means one has to
 go further back in time to see an equivalent horizon problem [58].
 There is also a causality problem, of sorts, as to why $c$ can
 change simultaneously over the whole universe and constantly stay equal
 throughout, once the value of $c$ has started to reduce and causal contact
 lost. This
 constraint is similar to the VT one for inflation in that the
 behaviour for $c$ really has to be pre-programed in the universe from
 its conception. Changing such constants also tends to  suppress any quantum
gravitational epoch at the beginning of the universe [59].
However, this quantum epoch can surface at a later time, which
must be pushed sufficiently far into the future [58]. In this
regard these model have some similarity with the pre-big bang
phase which also start in  a classical state and tend towards a
quantum gravitational singular region. This makes such models more
difficult to conceive of by quantum creation schemes, but it must
be admitted that quantization alone does not explain why creation
occurs. Neither, does quantum cosmology explain why the various
constants take their actual values, or why even the various
(quantum) laws of nature are applicable to the event.

 {\bf 8 Conclusions}

 We have seen how the constraint of VT for the initial patch size
 of an inflationary domain is actually the requirement that a
 singularity be present. The  horizon problem
  can still be solved by inflation for matter sources
 that initially have $\gamma<4/3$ before inflation occurs. For
 stiffer equations of state the requirement that the singularity
 be present means the particle horizon size is too small to keep
 the created inflationary domain within causal contact.
 Although, in practice the particle horizon size is likely to be
 modified by absorption and other quantum effects at high
 energies making the universe possibly opaque to light rays. The
 horizon problem could therefore be more difficult to allay than
 suggested by this classical analysis. Even having causal contact does
 not alone explain why the scalar field is homogeneous across the
 required domain size. This is, in some sense,
  a well ordered state, with low
 entropy that requires further explanation. The fact that the
 radiation case is discounted means the quantum vacuum state is
 necessarily unstable near the initial singularity [60]. The
 so-called Scalar Ricci curvature hypothesis (SRCH) cannot be
 implemented. It would be interesting to see if inflationary
 conditions are
 still compatible with  the Weyl-Curvature hypothesis cf.[5,60].

 The fact that inflation goes hand in hand with singularities is
 rather worrying. If singularities are anyway present how
 seriously should one take the inflationary paradigm. It would
 appear no better than the usual SBB model which also suffers from
 initial singularities so always emerging from unknown conditions.
 Even the violation of the strong energy condition is insufficient
 to remove  the singularity. One can postulate that the weak
 energy condition be violated to try and obviate this limitation,
  but this is rather drastic. It is likely to be very unstable,
   for example to
 perturbations and anisotropy increasing. Quantum gravitational
 effects might play a role since the proof that inflation has
 initial singularities depends on inflationary domains that finish
 inflating never again undergoing inflationary behaviour. If space
 could be recycled then the proofs of no past-eternal inflation
 might be obviated cf.[61]. If the Supernova data suggesting an
  accelerating universe[62]is confirmed then some
 possibility of inflation being re-entered  could be argued. It
 would certainly prove  that strong energy condition violating
 matter exists and that gravity can be  repulsive over large scales.
 Of course, even having an eternal model into the past does
  not entirely explain its
 existence. There is also the related suggestion that
  the presence of  closed timelike
 curves can give an explanation of the universe [34]. But time is an
 internal property of the universe and doesn't help  explain the coming
 into existence of the cosmological model.

 The other alternative to achieve inflation
 is to postulate that the inflationary stage
 is not preceded by an earlier phase but is started suddenly  by
 some ``quantum
 creation event''. As explained earlier, we are rather trying to
 use quantum reasoning to determine conditions assuming
 certain cosmological models
 are existing in the first place.  The event is rather analogous
  to a ``scanning tunneling microscope (STM) ''-see eg.[36],
   with the scalar potential $V(\phi)$ playing the role of the
   external potential applied to the ``tip of the probe''.
   This is rather uncertain extrapolation of quantum mechanics to
   a conceptually different type of problem.
   On a more practical level,
  even the notion that quantum tunneling takes place is dependent
  on the curvature dominating over matter at small scale factors.
  Firstly we don't know how curvature should be treated at quantum
  gravitational scales and secondly we have no reason to expect
  that matter should not be the dominant contribution. Although
  boundary conditions  impose such restrictions on what terms in
  the Wheeler DeWitt equation can dominate they arbitrarily remove
  generality from the approach. Unlike the (STM) example, in the
  cosmology case one is trying to determine the ``potential on the
  probe'' in order to predict the amount of inflation and it appears  a
  ``one shot'' event. To take the analogy further:
  we are trying to find the potential  while the probe itself is coming
  into existence.  This, at the very least,
  suffers from interpretational problems for making predictions.

Other related approaches to inflation such as the ending of the
pre-big bang phase or the variation of the speed of light have
related problems of causality. They require {\em coherence
lengths} that are larger than particle horizon lengths and so are
not possible to justify.

In conclusion, and in agreement with VT, inflation as a means of
resolving the horizon problem is difficult to justify: although
the possibility  is not entirely excluded for $\gamma<4/3$. The
consequence of having inflationary behaviour is that singularities
are present. A complete justification of having  inflationary
cosmological models will therefore require such singularities to
be removed or regularized in some way. Either by quantum
gravitational processes that violate the weak energy condition, or
circumscribed by quantum reasoning bypassing such singular
behaviour. This  still would not explain creation but would allow
us to take one step closer to formulating  the properties of the
event  that initially sparked the universe.

{\bf Acknowledgment}\\ I should like to thank D. Bak, C. Gordon,
A. Nesteruk, S.J. Rey and T. Vachaspati for helpful discussions.

\newpage

{\bf References}\\
\begin{enumerate}
\item T. Vachaspati and M. Trodden, Phys. Rev. D 61 (2000) p.
023502.\\ Mod. Phys. Lett. A 14 (1999) p.1661.
\item A.D. Linde, ``Particle Physics and Inflationary cosmology'',
(Harwood: Switzerland) (1990).
\item E.W. Kolb and M.S. Turner,
``The Early Universe'' (Addison-Wesley:New York) (1990).
\item S.W. Hawking and G.F.R. Ellis, ``The large scale structure
of the universe'' (Cambridge University press: Cambridge) (1973)
\item S.W. Hawking and R. Penrose, ``The nature of space and
time'' Princeton University press (1996)
\item W.G. Unruh, ``Is inflation the answer?'' in ``critical
dialogues in cosmology'' editor N. Turok, (World scientific press
: Singapore) (1997)
\item O.M. Moreschi, preprint gr-qc/9911105 (1999)
\item G.W. Gibbons, S.W. Hawking and J.M. Stewart, Nucl. Phys.
B 281 (1987) p.736.
\item S.W. Hawking and D.N. Page, Nucl. Phys. B 298 (1988) p.789.
\item D.N. Page, Phys. Rev. D 36 (1987) p.1607.\\
 D.H. Coule, Class. Quant. Grav. 12 (1995) p.455.
\item J.B. Hartle and S.W. Hawking, Phys. Rev. D 28 (1983) p.2960.
\item A. Vilenkin, Phys. Rev. D 30 (1984) p.549.\\
A.D. Linde, Sov. Phys. JEPT (1984) p.211.\\ V.A. Rubakov, Phys.
Lett. B 148 (1984) p.280.\\ Y.B. Zeldovich and A.A. Starobinski,
Sov. Astron. Lett.10  (1984) p.135.
\item A. Vilenkin, Phys. Rev. D 37 (1988) p.888.
\item D.N. Page, Phys. Rev. D 56 (1997) p.2065.
\item L. Randall and R. Sundrum, Phys. Rev. Lett. 83 (1999) p.
4690.
\item V.A. Belinsky, L.P. Grishchuk, Y.B. Zeldovich and I.M.
Klatatnikov, Sov. Phys.-JEPT 62 91985) p.195.\\ V.A. Belinsky and
I.M. Khalatnikov, Sov. Phys.-JEPT 66 (1987) p.441.
\item M.S. Madsen and P. Coles, Nucl. Phys. B 298 (1988) p. 2757.
\item H.J. Schmidt, Astron. Nachr. 311 (1990) p. 99.
\item D.S. Goldwirth and T. Piran, Phys. Reports 214 (1992) p.
223.
\item A.R. Liddle, A. Mazumdar and F.E. Schunck, Phys. Rev. D 58
(1998) p.061301.
\item P. Kanti and K.A. Olive, Phys. Lett. B 464 (1999) p.192.\\
Phys. Rev. D 60 (1999) p.043502.

\item H. Maris and S. Balibar, Physics Today, Feb. (2000) p. 29.
\item A. Linde, D. Linde and A. Mezhlumian, Phys. Rev. D 49 (1994)
p. 1783.
\item A. Borde and A. Vilenkin, ``The impossibility of
steady-state inflation'' Eighth Yukawa Symposium on relativistic
Cosmology Japan (1994).\\ Int. J. Mod. Phys. D 5 (1996) p.813.
\item P.R. Anderson, W. Eaker, S. Habib, C. Molina-Paris and E.
Mottola, preprint gr-qc/0005102
\item M.S. Volkov and A. Wipf, preprint hep-th/0003081 (2000)
\item R. Bousso, Phys. Rev. D 60 (1999) p.063503.
\item D.H. Coule and J. Martin, Phys. Rev. D 61 (2000) p.063501.
\item G.F.R. Ellis, A. Sumeruk, D. Coule and C. Hellaby,
Class. Quant. Grav. 9 (1992) p. 1535.\\ G.F.R. Ellis, Gen. Rel.
Grav. 24 (1992) p.1047.
\item J. Martin, Phys. Rev. D 49 (1994) p. 5086
\item F. Embacher, Phys. Rev. D 52 (1995) p.2150.
\item A. Carlini, D.H. Coule and D.M. Solomons, Mod. Phys. Lett. A
11 (1996) p.1453.
\item C. Schiller, preprint gr-qc/9610066.
\item J.R. Gott and Li-Xin Li, Phys. Rev. D 58 (1998) p. 023501.
\item  B.S. DeWitt, Phys. Rev. D 160 (1967) p.1113.\\
J.H. Kung, Gen. Rel. Grav. 27 (1995) p. 35.\\ M.L. Fil'chenkov,
Phys. Lett. B 354 (1995) p. 208.\\ H.C. Rosu and J. Socorro,
 Nuovo. Cim. B 113 (1998) p.119.
 \item R.W. Robinett, ``Quantum mechanics'', Oxford University
 Press: Oxford (1997)
 \item D.H. Coule, Mod. Phys. Lett. A 10 (1995) p.1989.
 \item R. Graham and P. Szepfalusy, Phy. Rev. D 42 (1990) p.2483.
\item M. Clutton-Brock, Quart. J. Roy. Astron. Soc. 34 (1993)
p.411.
\item J.W. Moffat, preprint gr-qc/9702014
\item P. Bak, ``How Nature works'', Springer-Verlag:New York
(1996).
\item A.H. Guth, S. Blau and E. Guendelman, Phys. Rev. D 35 (1987)
p.1747.\\ E. Fahri and A.H. Guth, Phys. Lett B 183 (1987 )
p.149.\\ E. Fahri, A.H. Guth and J. Guven, Nucl. Phys. B 339
(1990) p.417.
\item T. Padmanabhan,  ``Structure formation in the
universe'', Cambridge University press: Cambridge (1993) p.359.
\item J.M.M. Senovilla, Gen. Rel. Grav. 30 (1998) p.701
\item D. Bak and S.J. Rey, hep-th/9902173
\item R. Bousso, JHEP 9907 (1999) p.004.
\item N. Dadhich, J. Astrophysics. Astro. 18 (1997) p.343.
\item P. Binetruy, C. Deffayet, U. Ellwanger and D. Langlois,
Phys. Lett. B 477 (2000) p.285\\ E.E. Flanagan, S.H. Tye and I.
Wasserman, preprint hep-ph/9910498\\
 R. Maartens, D. Wands, B.A. Bassett and I. Heard, Phys. Rev. D 62
 (2000) p.041301.
\item D.J.H. Chung and K. Freese, preprint hep-th/9910235.
\item D. Hochberg and T.W. Kephart, Phys. Rev. Lett. 70 (1993)
p.2665.\\ L. Liu, F. Zhao and L.X. Li, Phys. Rev. D 52 (1995)
p.4752.
\item G. Veneziano, Phys. Lett. B 265 (1991) p.287.\\
M. Gasperini and G. Veneziano, Astropart. Phys. 1 (1993) p.317.
\item D.H. Coule, Class. Quant. Grav. 15 (1995) p.4752.\\
Phys. Lett. B 450 (1999) p.48.
\item S. Kawai, M. Sakagami and J. Soda, Phys. Lett. B 437 (1998)
p.284.\\ I. Antoniadis, J. Rizos and K. Tamvakis, Nucl. Phys. B
415 (1994) p.497.
\item L. Smolin, Class. Quant. Grav. 9 (1992) p.173.
\item J.W. Moffat, Int. J. of Mod. Phys. A 2 (1993) p.351.
\item A. Albrecht and J. Magueijo, Phys. Rev. D 59 p.043516.
\item J.D. Barrow,  Phys. Rev. D 59 (1999) p.043515.
\item D.H. Coule, Mod. Phys. Lett. A 14 (1999) p.2437.
\item J.D. Barrow and J. Magueijo, Class. Quant. Grav. 16 (1999) p.1435.
\item A.V. Nesteruk, Class. Quant. Grav. 11 (1994) p.L15.\\
Europhysics Lett. 36 (1996) p.233.
\item A.A. Starobinsky, preprint astro-ph/9912054.
\item S. Permutter et al. Nature 391 (1998) p.51 , see also preprint
astro-ph/9812133.\\ A.G. Riess et al. Astron. J. 116 (1998)
p.1009.
\end{enumerate}
\end{document}